\documentclass[11pt]{article}
\usepackage[letterpaper,hmargin=1in,vmargin=1.25in]{geometry}

\author{Hao Chen\\
Department of Computing and \\
Information Technology\\
Fudan University\\
Shanghai 200433,P.R.China\\
and\\
Jianhua Li\\
Department of Electronic Engineering\\
Shanghai JiaoTong University\\
Shanghai 200030, P.R.China }
\title{\bf Lower Bounds on \\
the Algebraic Immunity of Boolean Functions}
\date{May, 2006}

\begin{document}

\maketitle
\begin{abstract}
From the motivation of algebraic attacks to stream and block
ciphers([1,2,7,13,14,15]), the concept of {\em algebraic immunity}
(AI) of a Boolean function was introduced in [21] and studied in
[3,5,10,11,17,18,19,20,21]. High algebraic immunity is a necessary
condition for resisting algebraic attacks. In this paper, we give
some lower bounds on the algebraic immunity of Boolean functions.
The results are applied to give lower bounds on the AI of symmetric
Boolean functions and rotation symmetric Boolean functions. Some
balanced rotation symmetric Boolean functions with their AI near the
maximum
possible value $\lceil \frac{n}{2}\rceil$ are constructed. \\


\end{abstract}

{\bf I. Introduction and Preliminaries}\\

A Boolean function of $n$ variable is a mapping $f: F_2^n
\rightarrow F_2$, where $F_2$ is the field of two elements. The
weight of a Boolean function $wt(f)=|S_1(f)|$, where
$S_1(f)=\{(x_1,...,x_n): f(x_1,...,x_n)=1\} $ and $|*|$ is the
cardinality of the set. Any Boolean function has its algebraic normal form (ANF)\\
$$
\begin{array}{ccccccccccccc}
f(x_1,...,x_n)=a_0+\Sigma_{i_1<...<i_t} a_{i_1,...,i_t}
x_{i_1} \cdots x_{i_t}\\
\end{array}
$$
, where $a_0,..., a_{i_1,...,i_t}, \in F_2$. The (algebraic) degree
of $f$ is the number of  variables in the highest order term in the
above ANF. The Boolean function of degree 1 is called affine form.
Given a Boolean function $f$ of $n$ variables, a $n$ variable
Boolean function $g$ is called its annihilator function if $gf=0$,
or equivalently, $g$ is zero at all points of $S_1(f)$.  A Boolean
function is called balanced if the number of points in $S_1(f)$,
$wt(f)=2^{n-1}$. The distance of two Boolean functions $f$ and $g$
is $d(f,g)=|S_1(f-h)|$. The nonlinearity of a Boolean function $F$
is defined as $NL(f)=min_l \{d(f,l)\}$ where $l$ takes
over all possible affine forms (see [9]).\\

Boolean functions are widely used in block and stream ciphers, f.g.,
in S-boxes, combination generators and filter generators. It is
known that Boolean functions used in the practice of cryptography
have to satisfy some criteria, f.g., their degrees and
nonlinearities etc have to be high (see [9]). Algebraic attack was
proposed recently to block and stream ciphers (see
[1],[2],[7],[13],[14],[15]).   Because of some successful algebraic
attacks to several keystream generators, now it is interested to
understand the algebraic immunity $AI(f)$ of a Boolean function $f$,
which was introduced in [21]. General properties about algebraic
immunity of Boolean functions have been studied in
[3],[10],[11],[17],[19],[20],[21]. High algebraic immunity is a
necessary condition (but not sufficient) for resisting algebraic
attacks. It was proved that the AI of a $n$ variable Boolean
function is less than or equal to $\lceil \frac{n}{2}\rceil$ (see
[21]) . Recently several  algorithms for the computation for AI of
Boolean functions were given in[4]. If the $AI(f)$ of a Boolean
function $f$ is relatively small, the algorithms can be used to
determine the $AI(f)$ efficiently. However it is also known that
there are Boolean functions of $n$ variables with their $AI$ equal
to the maximal possible value $\lceil \frac{n}{2}\rceil$ (see
[5],[10],[12],[18]). Thus it is interesting to know more Boolean
functions with their AI equal to or near the
upper bound $\lceil \frac{n}{2}\rceil$.\\

A Boolean function is called symmetric if its value is determined by
the weight of its input vector. Symmetric Boolean functions have
been studied by many authors(see [8] and references there) from the
motivation of block and stream ciphers. In software and hardware
implementation the symmetric Boolean functions are efficient. Thus
it is interested to know the properties of AI of symmetric Boolean
functions.  In [5], the algebraic immunity of symmetric Boolean
functions was thoroughly studied. The AI of elementary symmetric
Boolean functions was explicitly determined and some symmetric
functions of maximum possible AI have been constructed. Rotation
symmetric Boolean functions (RSBF) were introduced and studied in
[22] for the purpose of fast hashing. A Boolean function $f$ on
$F_2^n$ is called rotation symmetric if
$f(x_1,x_2,...,x_n)=f(x_n,x_1,...,x_{n-1})$ for any
$(x_1,x_2,...,x_n)\in F_2^n$. The experimental studies of the
algebraic immunity of RSBF was initiated in [17]. From the
motivation of the possible use of symmetric and rotation symmetric
Boolean functions in cryptography , we are interested to have lower
bounds on the algebraic immunity of these functions and
the construction of these functions with relative high algebraic immunity.\\

We recall some basic facts about the algebraic immunity of a $n$
variable Boolean function( see [21],[10],[19],[3]).\\

{\bf Definition. } {\em Let $f$ be a Boolean function on $F_2^n$,
its algebraic immunity $AI(f)$ is defined to be the smallest number
$k$, such that,  there exists one Boolean function $g$ of degree $k$ which is
the annihilator function of $f$ or $1+f$.}\\

{\bf Theorem 1 (see [10],[21],[17]).} {\em Let $f$ be a $n$ variable
Boolean function. Then 1) $AI(f) \leq \lceil\frac{n}{2} \rceil$; 2)
$NL(f) \geq 2\Sigma_{i=0}^{AI(f)-2}C_{n-1}^i$, where $C_u^j$ is the
binomial coefficient; 3) If $AI(f) >d$ then $\Sigma_{i=0}^d
C_n^i \leq wt(f) \leq \Sigma _{i=0}^{n-(d+1)} C_n^i$.}\\

{\bf Theorem 2(see [3]).} {\em Let $f$ be a Boolean function of $n$
variables. Suppose  $wt(f) \geq 2^n-2^{n-d}$. Then any annihilator
of $f$ has its algebraic degree at least $d$.}\\

We note that Theorem 2 can not be applied directly to {\em balanced}
Boolean functions when lower bounding the AI of Boolean functions.
As far as our knowledge, there are quite few {\em explicitly given}
Boolean functions with the maximal possible AI and people do not
know much about how to lower bound the algebraic immunity of Boolean
functions (see [10],[12],[17],[18]). In this paper we apply Theorem
2 to the restrictions of Boolean functions on some affine subspaces
of $F_2^n$. Thus we present a method to obtain some lower bounds on
the algebraic immunity of Boolean functions. In this case, it is
possible that the restrictions of the annihilator functions on the
affine subspaces are zero. However if the affine subspaces are taken
sufficiently many, this consideration leads to
some useful results on the lower bound for the AI  of Boolean functions.\\

{\bf II. Main Result}\\

The following Theorem 3 is the main result of this paper.\\

{\bf Theorem 3.} {\em If $f$ is a Boolean function on $F_2^n$ and
$L_1$ (respectively $L_2$) is an affine subspaces with dimension $t$
(respectively $s$), such that , $|S_1(f|_{L_1})|>2^{t}-2^{t-d}$
(respectively $S_1((1+f)|_{L_2})|>2^s-2^{s-d}$). Then \\
1) either the annihilator functions of $f$ with minimum possible
degree (respectively the annihilator functions of $1+f$ with minimum
possible degree) have their degree at least $d$ or; \\
2) the annihilator functions of $f$ with minimum possible degree
(respectively the annihilator functions of $1+f$ with minimum
possible degree) are zero on $L_1$
(respectively on $L_2$).}\\

When Theorem 3 is applied to the balanced Boolean functions and
codimension $1$ affine subspace we have the following simple
conclusion. The proof of Corollary 1 is a direct application of Theorem 3.\\

{\bf Corollary 1.} {\em Let $f$ be a balanced Boolean function on
$F_2^n$ and $l$ is an affine form on $F_2^n$. Suppose $d(f,l) \geq
2^n-2^{n-d}$. Then we have, \\
1) either the algebraic immunity $AI(f)$ is at least $d$ or;\\
2) the annihilator functions of $f$ with the minimum possible degree
or the annihilator functions of $1+f$ with the minimum possible degree contain $l$ as a factor.}\\

In section III we can use Theorem 3 to give lower bounds on the
algebraic immunity of some symmetric and rotation symmetric
Boolean functions by using sufficiently many affine subspaces.\\

We also have the following result about the Hamming weight of the
restrictions of Boolean functions on affine subspaces.\\

{\bf Corollary 2.} {\em Let $f$ be a Boolean function on $F_2^n$
with $AI(f)=d+1$ and $L$ be a affine subspace of $F_2^n$ with
codimension $r$. Then the Hamming weight of $f$ restricted on $L$
satisfies $\Sigma_{i=0}^{d-r} C_{n-r}^{i} \leq wt(f|_{L}) \leq
\Sigma_{i=0}^{n-(d+1)} C_{n-r}^i$.}\\

When Corollary 2 applied to symmetric Boolean functions we have the
following result.\\

{\bf Corollary 3.} {\em Let $f$ be a $n$ variable symmetric Boolean
function. Then $f$ can not have the maximal possible algebraic
immunity $\lceil \frac{n}{2}\rceil$ in the following two cases.}\\
{\em 1) When $n$ is odd and $wt(x) \geq \lfloor  \frac{n}{2} \rfloor
$ , $f(x)$ is $1$ only when $wt(x)$ is odd (or only when $wt(x)$ is
even), $f(x)$ can be arbitrary for $wt(x) < \lfloor \frac{n}{2}\rfloor$.}\\
{\em 2) When $n$ is even and $wt(x) \geq \frac{n}{2}-1$, $f(x)$ is
$1$ only when $wt(x)$ is odd (or only when $wt(x)$ is even), $f(x)$
can be
arbitrary for $wt(x) < \frac{n}{2}-1$.}\\

By computing $d(f,l)$, where $l$ is the affine form $x_1+...+x_n$ or
$x_1+...+x_n+1$, and applying Corollary 2, we have the conclusion of
Corollary 3 immediately.\\

{\bf Proof of Theorem 3.} Let $g$ be an annihilator function of $f$,
that is $gf=0$. We have $(g|_{L_1})(f|_{L_1})=0$. From Theorem 2
$g|_{L_1}$ has its algebraic
degree at least $d$ if it is not a zero function. The conclusion is proved.\\

{\bf Proof of Corollary 2.} Let $l_1,...,l_r$ be $r$ linearly
independent affine forms such that  $L$ is defined by
$l_1=...=l_r=0$. Considering the Boolean function $f|_L$ as a
Boolean function of $n-r$ variables, if its algebraic immunity is
smaller $d-r$, we have a Boolean function $g'$ of $n-r$ variables
with algebraic degree at most $d-r$ such that $g'(f|_L)=0$ or
$g'((1+f)|_L)=0$. Thus the Boolean function $g=(l_1+1) \cdots
(l_r+1) g'$ can be think as a Boolean function of $n$ variables of
algebraic degree at most $d$. We have $gf=0$ or $g(1+f)=0$. This is
a contradiction. Therefore the algebraic immunity of $f|_L$ is at
least
$d-r+1$, we have the conclusion of 1) from the Theorem 1.\\

{\bf III. Lower Bound for AI of Symmetric and Rotation Symmetric
Boolean Functions}\\

In this section we use the main result to prove some lower bounds on
the  algebraic immunity of
symmetric and rotation symmetric Boolean functions.\\

{\bf A. Symmetric Boolean Functions}\\

{\bf Corollary 4.} {\em Let $f$ be a $n$ variable symmetric Boolean
function with simplified value vector
$v(f)=(v_0(f),...,v_i(f),...,v_n(f))$, i.e., $f(x)=v_i(f)$ when
$wt(x)=i$. Set\\
$$
\begin{array}{ccccccccc}
 U=\min \{\Sigma_{v_i(f)=1,i\leq \lceil n/2 \rceil} C_{\lceil
n/2 \rceil}^i, \Sigma_{v_i(f)=0,i\geq \lfloor n/2 \rfloor} C_{\lceil
n/2 \rceil}^{i-\lfloor n/2 \rfloor} \}
\end{array}
$$
Suppose $ U > 2^{\lceil n/2
\rceil }-2^{\lceil n/2 \rceil -d}$. Then $AI(f) \geq d+1$.}\\

{\bf Proof.} Let $i_1,...,i_{\lfloor \frac{n}{2} \rfloor}$ be arbitrary
 $\lfloor \frac{n}{2}\rfloor$ indices, $L_b$ be the dimension $\lceil \frac{n}{2}\rceil$ subspace
 of $F_2^n$ defined by $x_{i_1}=...=x_{i_{\lfloor \frac{n}{2}\rfloor}}=b$, where $b=0$ or $b=1$. If the
 condition of Corollary 4 is satisfied, $S_1(f|_{L_0})> 2^{\lceil n/2
\rceil }-2^{\lceil n/2 \rceil -d}$ and $S_1((1+f)|_{L_1})>2^{\lceil
n/2\rceil }-2^{\lceil n/2 \rceil -d}$. From Theorem 3, either
$AI(f)>d$ or the annihilator functions of $f$ or $1+f$ with minimum
possible degree are zero on $L_0$ and $L_1$. This implies that the
monomials in the algebraic normal forms $f$ (and $1+f$) have to
contain at least $\lceil \frac{n}{2}\rceil$ variables.
In the later case $AI(f)= \lceil \frac{n}{2}\rceil$. The conclusion is proved.\\

{\bf Example 1.} Let $f$ be a $15$ variable symmetric Boolean
function
$f=\sigma_2+\sigma_4+\sigma_6+\sigma_{10}+\sigma_{12}+\sigma_{14}$.
Then we have
its simplified value vector $v_f=(0,0,1,1,1,1,1,1,0,0,0,0,0,0,1,1)$. Then $U=246>240$ and $AI(f)\geq 5$\\

{\bf Example 2.} Let $f$ be a $n$ variable symmetric Boolean
function, $I=\{1,...,\lfloor \frac{n}{2} \rfloor, n-i\}-\{i\}$ where
$i \leq \lfloor \frac{n}{2} \rfloor$, $J=\{\lceil \frac{n}{2}
\rceil,...,n,i\}-\{n-i\}$. The symmetric Boolean function is defined
as follows.\\

$$
\begin{array}{ccccccc}
f(x)=1, wt(x) \in I\\
f(x)=0, wt(x) \in J\\
\end{array}
$$

Let $t$ be the smallest positive integer such that $C_{\lceil
\frac{n}{2} \rceil}^i+1 < 2^t$. It is clear $t<ilog_2n-i$.  We have
$U >2^{\lceil \frac{n}{2} \rceil}-2^t$ and $AI(f) \geq \lceil
\frac{n}{2} \rceil -t+1$. It is obvious that $t$ is asymptotically
less than $ilog_2 n$. These Boolean functions have their algebraic
immunities asymptotically
larger than $n/2- ilog_2 n+i-1$.\\

It is observed from Corollary 4 and Example 2, for a symmetric
Boolean function $f$ with the property that most vectors in $S_1(f)$
have their weight less than $\lceil \frac{n}{2}\rceil$ and most
vectors in $S_0(f)$ have their weight larger than $\lceil
\frac{n}{2}\rceil$, its AI is relatively high. This suggests that
 these symmetric Boolean functions can be
possibly used in stream ciphers, if they satisfy other cryptographic
criteria.\\

{\bf B.  Rotation Symmetric Boolean Functions}\\

In this subsection we use Theorem 3 to give lower bound
for the algebraic immunity of RSBFs.\\

{\bf Example 3.} Let $f$ be a rotation symmetric Boolean function
of $6$ variable\\
$$
\begin{array}{cccccccccc}
f=x_1x_2x_3+x_2x_3x_4+x_3x_4x_5+x_4x_5x_6+x_5x_6x_1+x_6x_1x_2\\
+x_1x_4+x_2x_5+x_3x_6+x_1x_3x_5+x_2x_4x_6+\\
x_1x_2x_3x_4+x_2x_3x_4x_5+x_3x_4x_6x_1+\\
x_1x_2x_3x_4x_5+x_2x_3x_4x_5x_6+x_3x_4x_5x_6x_1+x_4x_5x_6x_1x_2+x_5x_6x_1x_2x_3+x_6x_1x_2x_3x_4
\end{array}
$$

This is a balanced Boolean function with nonlinearity $24$ and $\Delta(f)=40$, which
satisfies $PC(2)$ criteria (see [24]).\\

We consider two affine subspaces $L_1$ (respectively $L_2$) in
$F_2^6$ defined by $x_1=x_2=x_3=0$(respectively $x_1=1,x_2=x_3=0$).
It is easy to check that $S_1((1+f)|_{L_1})$ has $7$ points (in
$L_1$) and $S_1(f|_{L_2})$ has $5$ points( in $L_2$). Thus the
annihilator functions of $1+f$ (respectively, $f$) have degree at
least $2$ or are zero on $L_1$ (respectively $L_2$). In the later
case, the annihilator functions of $1+f$ (respectively, $f$) are
zero on any rotation transformation of $L_1$ (respectively, $L_2$).
From this observation, we have $AI(f) \geq
2$.\\

{\bf Example 4.} It is clear that each orbit in $F_2^n$ under the
circular action $\rho (x_1,x_2,...,x_n)=(x_n,x_1,...,x_{n-1})$
contains $h$  elements, where $h$ is a factor of $n$. On the other
hand the orbit of a weight $i$ vector in $F_2^n$ under the action of
all permutations contains $C_n^i$ elements, which is the union of
orbits of circular actions.\\

From [5] and [8] we know the following {\em Balanced} symmetric
Boolean function $f$ of $n$ ($n$ is odd) variables
has the maximal possible AI $\lceil \frac{n}{2} \rceil$.\\
$$
\begin{array}{cccccc}
f(x)=1, wt(x) <\lceil \frac{n}{2} \rceil\\
f(x)=0, wt(x)  \geq \lceil \frac{n}{2} \rceil
\end{array}
$$
When $n$ is even,the value $b$ in the following definition can be
suitably chosen
 such that it is balanced(in this case the function is not symmetric, however it can be rotation symmetric if
 $b$ is chosen to be the same on the orbits of circular actions).\\
$$
\begin{array}{cccccc}
f(x)=1, wt(x) < \frac{n}{2} \\
f(x)=0, wt(x) <  \frac{n}{2} \\
f(x)=b \in F_2, wt(x)=\frac{n}{2}
\end{array}
$$

If we exchange some orbits under circular actions in the  two sets
$S_0(f)$ and $S_1(f)$,  we get some rotation symmetric Boolean
functions and the lower bound on their $AI$ can be proved by
applying Theorem 3. Let $H \subset S_0(f)$ and $H' \subset S_1(f)$
be two subsets with the same cardinality , which are the union of
orbits under circular actions. Set $X=S_0(f)\bigcup
H'-H,X'=S_1(f)\bigcup H-H'$. Let $f'$ be the Boolean function with
$S_0(f')=X, S_1(f')=X'$. This is a balanced
Boolean function. We have the following result.\\

{\bf Corollary 5.} {\em $AI(f') > \lceil \frac{n} {2} \rceil -\lceil
log_2 |H| \rceil$.}\\

When $n$ goes to infinity, we have constructed  some balanced
rotation symmetric Boolean functions  with their algebraic immunity
asymptotically equal to $ \lceil \frac{n}{2} \rceil
-log_2 n $ if $|H|=|H'|=n$ (f.g., $H$ and $H'$ consist of  one orbit).\\

{\bf Proof.} Let $i_1,...,i_{\lfloor \frac{n}{2} \rfloor}$ be
arbitrary  $\lfloor \frac{n}{2}\rfloor$ distinct indices, $L_b$ be
the dimension $\lceil \frac{n}{2}\rceil$ subspace
 of $F_2^n$ defined by $x_{i_1}=...=x_{i_{\lfloor \frac{n}{2}\rfloor}}=b$, where $b=0$ or $b=1$. We have
  $S_1(f')\supset S_1(f)-H'$ and $S_1(f'|_{L_0})> 2^{\lceil n/2 \rceil }-2^{d}$, where $d= \lceil
log_2 |H| \rceil$. Similarly we have $S_1(1+f') \supset S_1(1+f)-H$
and  $S_1((1+f')|_{L_1})>2^{\lceil n/2\rceil }-2^{d}$. From Theorem
3, either $AI(f)>\lceil \frac{n} {2} \rceil -\lceil log_2 |H|
\rceil$ or the annihilator functions of $f'$ or $1+f'$ are zero on
$L_0$ and $L_1$. This implies that the monomials in the algebraic
normal forms $f'$ and $1+f'$ have to contain at least $\lceil
\frac{n}{2}\rceil$
variables. In the later case $AI(f)= \lceil \frac{n}{2}\rceil$. The conclusion is proved.\\

{\bf IV. Conclusion}\\

We presented a method to obtain some lower bounds on the algebraic
immunity for Boolean functions. When the results are applied to
symmetric or rotation symmetric Boolean functions, some  lower
bounds on the  algebraic immunity can be proved for these Boolean
functions. Some rotation symmetric Boolean functions with their AI
near the maximal possible value $\lceil\frac{n}{2}\rceil$ are
constructed. Our method suggested some symmetric and rotation
symmetric Boolean functions of large number of variables with high
algebraic immunity. Thus they can be possibly  used in stream
ciphers
if these Boolean functions satisfy other cryptographic criteria. \\

{\bf Acknowledgement.} The work of the 1st author's was supported in
part by NNSF of China under Grant 90607005 and Distinguish Young
Scholar Grant 10225106.\\

\begin{center}
REFERENCES
\end{center}

[1]F.Armknecht and M.Krause, Algebraic attacks on stream combiners
with memory, in Advances in Cryptology-Crypto2003, LNCS 2729, pages 162-176, Springer-Verlag.\\

[2] F.Armknecht, Improving fast algebraic attacks, in Fast Software
Encryption -2004, LNCS 3017, pages 65-82, Springer-Verlag.\\

[3] F.Armknecht,On the existence of low-degree equations for
algebraic attacks, Cryptology e-print Archive, 2004/185\\

[4] F.Armknecht, C.Carlet, P.Gaborit, S.Kunzli, W.Meier and
O.Ruatta, Efficient computation of algebraic immunity for algebraic
and fast algebraic attacks, Advances in Cryptology -Eurocrypt 2006,
LNCS 4004, pages 147-164.\\

[5] An Braeken and B.Preneel, On the algebraic immunity of
symmetric Boolean functions, Indocrypt 2005.\\

[6] An Braeken, J.Lano and B.Preneel, Evaluating the resistance of
stream ciphers with linear feedback against fast algebraic attacks,
ACISP 2006, LNCS 4058, pages 40-51.\\

[7] A.Canteaut, Open problems related to algebraic attacks on
stream ciphers, In WCC 2005, pages 1-10.\\

[8] A.Canteaut and M.Videau, Symmetric Boolean functions, IEEE
Transactions on Information theory, vol. 51(2005), no. 8, pages
2791-2811.\\

[9] C. Carlet  "Boolean Functions for Cryptography and Error
Correcting Codes" (150 pages), chapter of the monography  ``Boolean
methods and models" published by Cambridge University Press (Peter
Hammer et Yves Crama editors).\\

[10] C.Carlet, D.K.Dalai, K.C.Gupta and S.Maitra, Algebraic immunity
for crypotographically significant Boolean functions: analysis and
construction, IEEE Trans. Inf. Theory, vol.52(2006), no.7, pages
3105-3121.\\

[11] C.Carlet, On the Higher Order Nonlinearities of Algebraic
Immune Functions, Advances in Cryptology-Crypto 2006, LNCS 4117.\\

[12] C.Carlet, A method of construction of balanced functions with
optimum algebraic immunity, Cryptology e-print Archive, 2006\\

[13] N.Courtois and W.Meier, Algebraic attacks on stream ciphers
with linear feedback,
in Advances in Cryptology-Eurocrypt 2003, LNCS2656, pages 346-359, Springer-Verlag.\\

[14] N.Courtois, Fast algebraic attacks on stream ciphers with
linear feedback,
in Advances in Cryptology-Crypto2003, LNCS 2729, pages 176-194, Springer-Verlag.\\

[15] N.Courtois and J.Pieprzyk, Cryptanalysis of block ciphers with
overdetermined systems of equations, in Advances in
Cryptology-Asiacrypt2002, LNCS, 2501, pages 267-287.\\

[16] T.W.Cusick and P. Stanica, Fast evaluation, weighted and
nonlinearity of rotation-symmetric functions, Discrete Math.,
vol.258(2002), pages 289-301.\\

[17] D.K.Dalai,K.C.Gupta and S.Maitra, Results on algebraic immunity
of cryptographically significant Boolean functions, in Indocrypt 2004, LNCS 3348\\

[18] D.K.Dalai, S.Maitra and S.Sarkar, Basic Theory in construction
of Boolean functions with maximal possible annihilator immunity,
Cryptology e-print Archive, 2005/229.\\

[19] J.D.Golic, Vectorial Boolean functions and induced algebraic
equations, IEEE Transactions on Information Theory, vol. 52,
no. 2, pages 528-537, Feb.2006.\\

[20] G. Gong, On existence and invariant of algebraic attacks,
preprint.\\

[21] W.Meier, E.Pasalic and C.Carlet, Algebraic attacks and
decomposition of Boolean functions, in Advances in
Cryptology-Eurocrypt-2004, LNCS 3027, pages 474-491,
Springer-Verlag.\\

[22] J. Pieprzyk and C.X.Qu, Fast hashing and rotation symmetric
functions, J.Universal Comput.Sci., vol.5(1999), pages 20-31.\\

[23] P.Stanica and S.Maitra, A constructive count of rotation
symmetric functions, Information processing Letter, vol. 88(2003),
pages 299-304.\\

[24] P.Stanica and S.Maitra, Rotation symmetric Boolean functions
-Count and cryptographic
properties, preprint.\\

\end{document}